\documentstyle[prl,aps]{revtex}
\begin{document}
\title {A Piecewise-Conserved Constant of Motion for a Dissipative System}
\author {Lior M. Burko} 
\address{Theoretical Astrophysics 130-33, 
California Institute of Technology, Pasadena, CA 91125}
\date{\today}
\maketitle

\begin{abstract}
We discuss a piecewise-conserved constant of motion for a simple 
dissipative oscillatory mechanical system. The system is a harmonic
oscillator with sliding (dry) friction.  
The piecewise-conserved constant of motion corresponds to 
the time average of the energy of the system during one 
half-period of the motion, and changes abruptly at the turning points of 
the motion. At late times the piecewise-conserved constant of motion 
degenerates to a constant of motion in the usual sense.
\end{abstract}

\section{Introduction}
Finding constants of motion is an important step in the solution of many 
problems of Physics, as they allow to reduce the number of the degrees of 
freedom of the problem. Constants of motion are intimately related to 
conservation laws or symmetries of the system. For example, it is well 
known that a symmetry of the Lagrangian of a system is responsible, by 
virtue of Noether's theorem, to a constant of motion \cite{goldstein}. 
By definition, a constant of motion preserves its 
value during the evolution of the system. Even in cases where there are 
no constants of motion, one could still 
sometimes find adiabatic invariants, 
which generalise the concept of a constant of motion to 
systems with slowly-varying parameters. It turns out, however, that there 
are cases where there are constants of motion, which are only 
piecewise-conserved. 
Well-known examples of such piecewise-conserved constants of motion 
arise from the generalisation of the Laplace-Runge-Lenz vector for 
general central potentials. For example, in the case of the 
three-dimensional isotropic harmonic oscillator, the Fradkin vector 
\cite{fradkin-67} 
(which generalises the Laplace-Runge-Lenz vector) abruptly reverses its 
direction (although preserves its magnitude) during a full period 
\cite{buch-75} (see also \cite{heintz}). 
(The Fradkin vector directs toward the perigee, and the position of the 
perigee jumps discontinuously whenever the particle passes through the 
apogee.) Also, in the truncated Kepler problem, the 
Peres-Serebrennikov-Shabad vector abruptly changes its direction whenever 
the particle in motion passes through the periastron 
\cite{peres-79,serebrennikov-71}. Again, it is just 
the direction of the conserved vector which is only piecewise-conserved: 
the magnitude of the vector remains a constant of motion in the original 
meaning (namely, the magnitude has a fixed value throughout the motion). 
Piecewise-conserved constants of motion may also be 
relevant for systems which involve radiation reaction. 
In the above examples, 
the piecewise-conserved constants of motion appear in non-dissipative 
systems, and result from a discontinuity of the force (as in the 
truncated Kepler problem) or from geometrical considerations (as in the 
three-dimensional isotropic harmonic oscillator case). These  
piecewise-conserved constants of motion involve 
vectors rather than scalars, which still conserve their magnitude. 

The following question arises: Can one find, for elementary systems,  
piecewise-conserved constants of motion? 
In what follows, we shall discuss an 
elementary piecewise-conserved scalar constant of motion, for a simple 
oscillatory mechanical 
model which involves dissipation in the form of sliding (dry) friction. 
Although dry friction is more nearly descriptive of everyday macroscopic
motion in inviscid media, it is usually ignored in elementary mechanics
courses and textbooks, which very frequently discuss viscous friction
(which is velocity dependent). Dry friction  exhibits, however, some very
interesting features and can be readily presented in the laboratory or
classroom demonstration. 

The problem of a harmonic oscillator damped by dry friction was considered
by several authors. Lapidus analised this problem for equal coefficients
of static and kinetic friction, and found the position where the
oscillator comes to rest \cite{lapidus-70}. Hudson and Finfgeld were able
to find the general solution of the equation of motion
\cite{hudson-finfgeld71}. However, they again assumed equal coefficients
of static and kinetic friction, and used the Laplace transform technique,
which is unknown to students of elementary mechanics courses, to generate
the solution. An elementary solution, which
ignores static friction, was derived by Barratt and Strobel
\cite{barratt-strobel81}.  This solution is based on solving separately
for
each half cycle of the motion, and is consequently tedious and
unappealing. Recently, Zonetti {\it et al.} \cite{zonetti-99} considered
the
related problem of both dry and viscous friction for a pendulum, but did
not offer a full analytic solution for the motion. 

In this work, we find the general solution for the motion taking into
account both static and kinetic friction, using elementary techniques
which are available for students of elementary courses of mechanics. We
analyse the solution using a piece-wise conserved constant of motion.   
The discussion, as well as the corresponding laboratory experiment of
classroom demonstration, are suitable for a basic course for physics or
engineering students.

\section{Elementary Discussion}
Let a block of mass $M$ be placed on a horizontal surface, such that the 
coefficients of static and kinetic friction are $\mu_s$ and $\mu_k$, 
respectively.  
The block is attached to a linear spring with spring 
constant $k$ (for both compression and extension), such that initially the
spring is 
stretched from its equilibrium length by $\ell$, and the block is kept at 
rest at $x=\ell$. At time $t=0$ the block is released. If $k\ell > \mu_sMg$, 
$g$ being the gravitational acceleration, the block would start to 
accelerate. We take the friction to be small, namely $\mu_{s,k} \ll 
\ell k/(Mg)$, and also assume slow motion, such that the friction force 
is independent of the speed. Namely, we neglect effects such as 
air-resistance, and include only the force which results from the block 
touching the surface. We also neglect any variation of $\mu_k$ with the 
speed.  

Immediately after the block 
starts accelerating, its motion is governed by the equation of motion 
\begin{equation}
M\ddot{x}=-kx+\mu_kMg, 
\label{eq1}
\end{equation}
with initial conditions $x(0)=\ell$ and $\dot{x}(0)=0$. From now on, let 
us introduce the frequency $\omega^2=k/M$. Of course, the system does not 
preserve its energy, due to the friction force. However, let us define 
a new coordinate $x'=x-\mu_kg/\omega^2$. Equation (\ref{eq1}) then 
becomes $\ddot{x'}+\omega^2x'=0$. For this equation we know that there is 
a constant of motion, namely 
${\cal E}=\frac{1}{2}M\dot{x'}^2+\frac{1}{2}M\omega^2x'^2$. Therefore, 
despite the presence of friction, one can still find a constant of 
motion, which has the functional form of the total mechanical energy, but 
which is of course {\em not} the energy, as the latter is not conserved. 
Calculating its numerical value we find that 
${\cal E}_0=\frac{1}{2}M\omega^2\left(\ell-\mu_kg/\omega^2\right)^2$.

At the time 
$t=\pi/\omega$ the velocity of the block vanishes, and it can be easily 
shown that at $t=\pi/\omega$ its acceleration is 
$\ddot{x}(\pi/\omega)=(\ell-\mu_kg/\omega^2)\omega^2>0$, such that the 
block reverses its motion. (We assume here that 
$M\ddot{x}(\pi/\omega)>\mu_sMg$.) The nature of the friction force is 
that its direction is always opposite to the direction of motion. 
Consequently, the equation of motion now changes to 
\begin{equation}
M\ddot{x}=-kx-\mu_kMg,
\label{eq2}
\end{equation}
with initial conditions $x(\pi/\omega)=\ell-2\mu_kg/\omega^2$ and 
$\dot{x}(\pi/\omega)=0$. One can again solve this equation readily. This 
time, let us define $x'=x+\mu_kg/\omega^2$. Equation (\ref{eq2}) again 
becomes $\ddot{x'}+\omega^2x'=0$, such that ${\cal E}$ is still 
conserved. However, this time the {\em numerical value} of ${\cal E}$, 
which we denote by ${\cal E}_1\ne{\cal E}_0$, and we find 
${\cal E}_1=\frac{1}{2}M\omega^2\left(\ell-3\mu_kg/\omega^2\right)^2$. 
One can 
describe the next phases of the motion similarly. During each phase of 
the motion (during half a period between two times at which the velocity 
vanishes) ${\cal E}$, if defined properly, is conserved. However, ${\cal 
E}$ is only piecewise-conserved, as its value changes abruptly from phase 
to phase. We note that the period 
of the oscillations is not altered by the presence of friction, and 
denote by $P_{1/2}$ half that period. Namely, $P_{1/2}\equiv \pi/\omega$.

\section{General Discussion}

Let us now discuss the system in a more general way. It turns out that 
although there are friction forces, one can still write a hamiltonian 
$$H(x,p;t)=\frac{1}{2M}p^2+\frac{1}{2}M\omega^2 x^2-f(t)x,$$ 
where $f(t)=(-1)^{\left[t/P_{1/2}\right]}\mu_kMg$.
The equation of motion is now
\begin{equation}
\ddot{x}+\omega^2 x=f(t)/M
\label{eq3}   
\end{equation}
with the initial 
conditions being (as before) $x(0)=\ell$ and $\dot{x}(0)=0$. 
We denote by 
square brackets of some argument the largest integer smaller than or 
equal to the argument. We also assume that the 
static friction force at the turning points of the motion is smaller than 
the elastic force of the spring, such that the motion does not stop. (Of 
course, for large enough time, this would not be true any more, and the 
block would eventually stop---see below.) Let us define the (complex) 
variable 
$\xi\equiv \dot{x}+i\omega x$ \cite{landau}, where $i^2=-1$. Then, instead 
of a real second 
order equation (such as Eq. (\ref{eq3})), one obtains a complex first 
order equation. It is advantageous to do this, because there is a general 
solution for any inhomogeneous linear first-order differential equation 
in terms of quadratures. 
Substituting the definition for $\xi$ in Eq. (\ref{eq3}) we find that 
the equation of motion, in terms of $\xi$, takes the form 
\begin{equation}
\dot{\xi}-i\omega\xi=f(t)/M,
\label{eq4}
\end{equation}
with the initial condition $\xi_0\equiv \xi(t=0)=i\omega\ell$. 
The solution of the equation of motion (\ref{eq4}) is 
\begin{equation}
\xi (t)=e^{i\omega t}\left\{\int_{0}^{t}\frac{1}{M}f(t')e^{-i\omega 
t'}\,dt'+\xi_0\right\}.
\label{eq5}
\end{equation}
After finding the solution $\xi(t)$ we can find $x(t)$ and $\dot{x}(t)$ 
by $\dot{x}(t)=(\xi(t)+\xi ^{*}(t))/2$ and 
$x(t)=(\xi(t) -\xi ^{*}(t))/(2i)$. We denote by a star complex 
conjugation. In order to integrate Eq. (\ref{eq5}) we find it convenient to 
separate the discussion to two cases: case (a) where 
$\left[t/P_{1/2}\right]$ is an odd number (namely, 
$\left[t/P_{1/2}\right]=2n-1$), 
and case (b) where $\left[t/P_{1/2}\right]$ is even (namely, 
$\left[t/P_{1/2}\right]=2n$), where $n$ is integer.  We next split the 
interval of integration in Eq. (\ref{eq5}) into two parts: we first 
integrate from $t'=0$ until $t_{2n-1}\equiv(2n-1)P_{1/2}$, 
and then 
integrate from $t_{2n-1}$ to $t$, and sum the two 
contributions. 
Integrating term by term we find
\begin{eqnarray}
\int_0^{t_{2n-1}}\frac{1}{M}f(t')e^{-i\omega 
t'}\,dt'&=&\sum_{j=0}^{n-1}\int_{2jP_{1/2}}^{(2j+1)P_{1/2}}\mu_k 
ge^{-i\omega 
t'}\,dt' \nonumber \\
&-&\sum_{j=0}^{n-2}\int_{(2j+1)P_{1/2}}^{(2j+2)P_{1/2}}\mu_k 
ge^{-i\omega t'}\,dt' 
\nonumber \\ &=&-2i(2n-1)\mu_kg/\omega.
\end{eqnarray}
For case (a) we find that 
\begin{equation}
\int_{t_{2n-1}}^t\frac{1}{M}f(t')e^{-i\omega
t'}\,dt'=-\int_{t_{2n-1}}^t\mu_kge^{-i\omega
t'}\,dt'=-i(e^{-i\omega t}+1)\mu_kg/\omega.
\end{equation} 
For case (b) we find
\begin{eqnarray}
\int_{t_{2n-1}}^t\frac{1}{M}f(t')e^{-i\omega
t'}\,dt'&=&-\int_{(2n-1)P_{1/2}}^{2nP_{1/2}}\mu_kge^{-i\omega
t'}\,dt'+\int_{2nP_{1/2}}^{t}\mu_kge^{-i\omega
t'}\,dt'\nonumber \\&=&i(e^{-i\omega t}-3)\mu_kg/\omega.
\end{eqnarray}
Collecting the two integrals, we find for case (a) that 
\begin{equation}
\xi_a(t)=-2i(2n-1)e^{i\omega t}\mu_kg/\omega-i(1+e^{i\omega 
t})\mu_kg/\omega +i\omega\ell e^{i\omega t},
\end{equation} 
and for case (b)
\begin{equation}
\xi_b(t)=-2i(2n-1)e^{i\omega t}\mu_kg/\omega+i(1-3e^{i\omega 
t})\mu_kg/\omega +i\omega\ell e^{i\omega t}.
\end{equation}
Recalling the different values of $\left[t/P_{1/2}\right]$ for the two 
cases (a) and (b), we can unify the expressions for both $\xi_a(t)$ and 
$\xi_b(t)$, namely
\begin{equation}
\xi(t)=-i\left(2\left[t/P_{1/2}\right]+1\right)e^{i\omega 
t}\mu_kg/\omega+(-1)^{\left[t/P_{1/2}\right]}i\mu_kg/\omega 
+i\omega\ell e^{i\omega t}.
\end{equation}
From this solution for $\xi(t)$ we can find that 
\begin{equation}
x(t)=(-1)^{\left[t/P_{1/2}\right]}\mu_kg/\omega^2+
\left\{\ell-\left(2\left[t/P_{1/2}\right]+1\right)\mu_kg/\omega^2\right\}
\cos\omega t,
\label{pos}
\end{equation}
and
\begin{equation}
\dot{x}(t)=-\left\{\ell-\left(2\left[t/P_{1/2}\right]+1\right)
\mu_kg/\omega^2\right\}
\omega\sin\omega t.
\label{vel}
\end{equation}
An interesting property of the solution given by Eqs. (\ref{pos}) and 
(\ref{vel}) is that for each half cycle it looks as if the motion were 
that of a simple harmonic oscillator, with no friction. In fact, the 
effect of the friction for each half cycle enters only in the initial 
conditions for that half cycle, or, more accurately, in the smaller value 
for the initial position for the half cycle. In addition, it is evident 
from Eq. (\ref{pos}) that the damping of the amplitude of the oscillation 
is linear in the time $t$, whereas in disspative systems in which the 
resistance is speed-dependent the damping is exponential in the time. 

Let us now define a new coordinate 
$x'(t)\equiv 
x(t)-f(t)/(M\omega^2)=x(t)-(-1)^{\left[t/P_{1/2}\right]}\mu_kg/\omega^2$. 
Then, we find that 
\begin{equation}
x'(t)=\left\{\ell-\left(2\left[t/P_{1/2}\right]+1\right)\mu_kg/\omega^2\right\}
\cos\omega t,
\end{equation}
and
\begin{equation}
\dot{x'}(t)=-\left\{\ell-\left(2\left[t/P_{1/2}\right]+1\right)
\mu_kg/\omega^2\right\}
\omega\sin\omega t.
\end{equation}
Next, we define  
\begin{equation}
{\cal E}(t)=\frac{1}{2}M\dot{x'}^2(t)+\frac{1}{2}M\omega^2{x'}^2(t).
\end{equation} 
Substituting the expressions for $x'(t)$ and $\dot{x'}(t)$ in ${\cal E}$, 
we find that 
\begin{equation}
{\cal 
E}(t)=\frac{1}{2}M\omega^2\left\{\ell-\left(2\left[t/P_{1/2}\right]+1\right)
\mu_kg/\omega^2\right\}^2.
\end{equation}
It is clear that ${\cal E}$ is not a constant of motion. However, a close 
examination shows that it is piecewise conserved: the only dependence on 
$t$ is through $\left[t/P_{1/2}\right]$. Therefore, between any two 
consecutive turning points we find that the numerical value of ${\cal E}$ 
is conserved. Consequently, ${\cal E}$ is a piecewise-conserved constant 
of motion. 

Clearly, ${\cal E}$ has the dimensions of energy. However, we 
stress that ${\cal E}$ is not the mechanical energy of system, because 
the latter is not even piecewise-conserved. In fact, the total mechanical 
energy of the system is
$T(t)=\frac{1}{2}M\dot{x}^2(t)+\frac{1}{2}M\omega^2{x}^2(t)$, namely, 
\begin{eqnarray}
T(t)&=&
\frac{1}{2}M\omega^2\left\{\ell-\left(2\left[t/P_{1/2}\right]+1\right)
\mu_kg/\omega^2\right\}^2+\frac{1}{2}M\frac{{\mu_k}^2g^2}{\omega^2} 
\nonumber \\
&+&(-1)^{\left[t/P_{1/2}\right]}\mu_kMg
\left\{\ell-\left(2\left[t/P_{1/2}\right]+1\right)
\mu_kg/\omega^2\right\}\cos\omega t,
\label{energy}
\end{eqnarray} 
which is a monotonically decreasing function of $t$, as expected. (Notice 
that whenever the cosine changes its sign, so does its amplitude.) Of 
course, if we add to $T(t)$ the work done by the friction force, we 
obtain a constant value. The fact that the total mechanical energy is 
monotonically decreasing is important: the system loses energy 
constantly. We have previously noted that the position and the velocity 
of the block during each half cycle are influenced by the presence of 
friction only through the initial conditions for that half cycle, but 
otherwise the motion is simple oscillatory. Despite this fact, the loss 
of energy occurs throughout of motion, as is evident from Eq. 
(\ref{energy}), as should be expected. 

In order to gain some more insight into the 
meaning of the piecewise-conserved ${\cal E}$, let us find the time 
average of $T(t)$ between two successive turning points. Clearly, the 
average of the cosine vanishes, and we find 
\begin{eqnarray}
<T(t)>&=&\frac{1}{2}M\omega^2\left\{\ell-\left(2\left[t/P_{1/2}\right]+1\right)
\mu_kg/\omega^2\right\}^2+\frac{1}{2}M\frac{{\mu_k}^2g^2}{\omega^2}
\nonumber \\ &=&{\cal 
E}(t)+\frac{1}{2}M\frac{{\mu_k}^2g^2}{\omega^2}.
\end{eqnarray}
Therefore, the physical meaning of ${\cal E}$ is the following: up to a 
global additive constant (namely, a constant throughout the motion) 
${\cal E}$ is equal to the time average of the total mechanical energy of 
the system $T(t)$ between any two consecutive turning points. Because of 
the dissipation, this time average decreases from one phase of the motion 
to the next, and therefore ${\cal E}$ is only piecewise conserved. 

\begin{figure}
\input epsf
\centerline{ \epsfxsize 8.5cm
\epsfbox{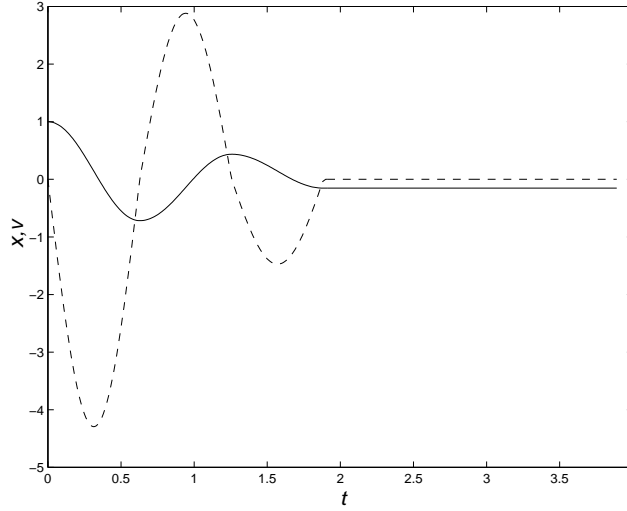}}
\caption{The position and the velocity as functions of time $t$, for the
choice of parameters $\mu_s=0.54$ and $\mu_k=0.36$. Solid
line: $x(t)$. Dashed line: $\dot{x}(t)$.}
\label{fig1} \end{figure}

\begin{figure}
\input epsf
\centerline{ \epsfxsize 8.5cm
\epsfbox{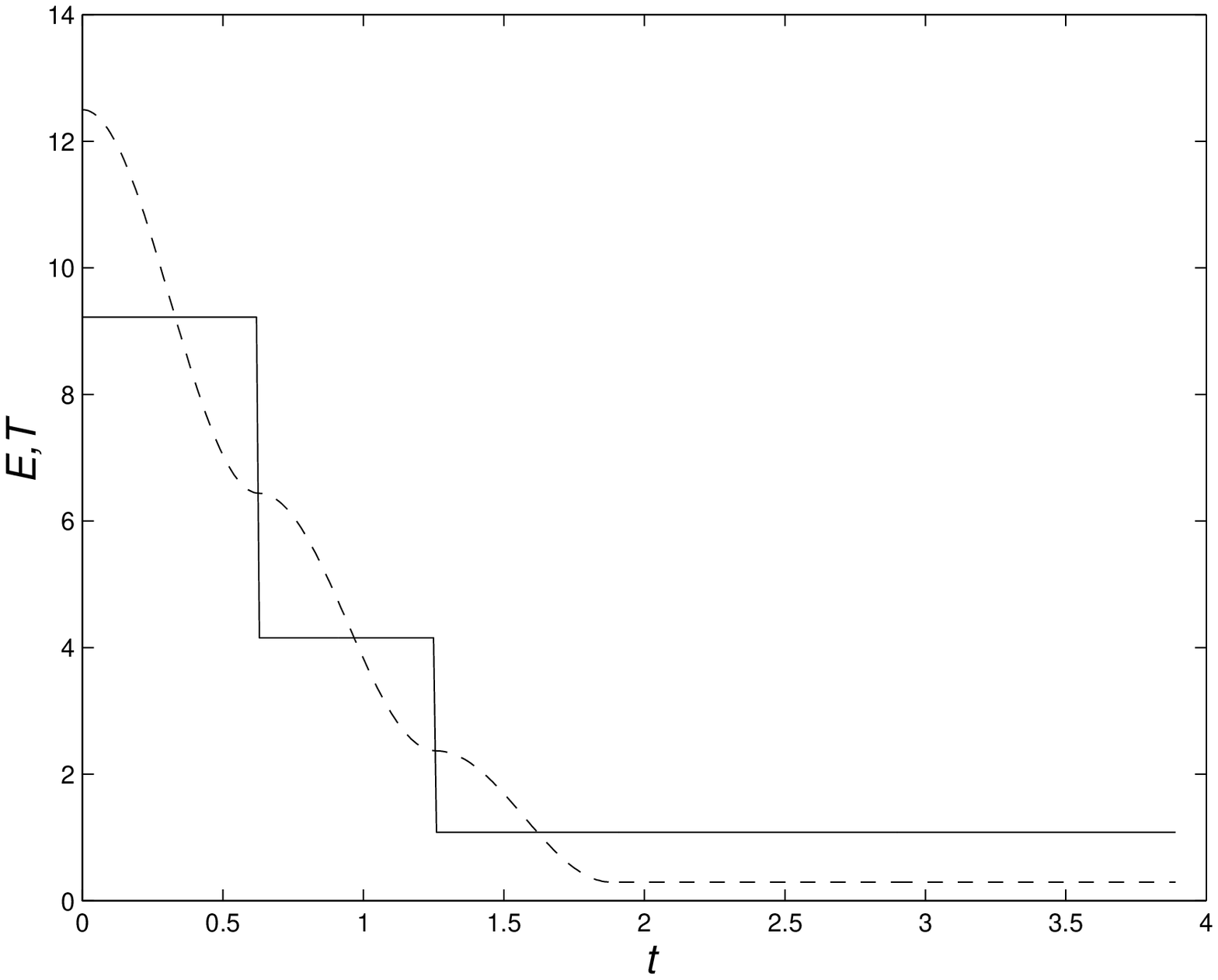}}
\caption{The picewise-conserved constant of motion and the total
mechanical energy as functions of time $t$, for the
choice of parameters $\mu_s=0.54$ and $\mu_k=0.36$. Solid
line: ${\cal E}(t)$. Dashed line: $T(t)$.}
\label{fig2} \end{figure}

\begin{figure}
\input epsf
\centerline{ \epsfxsize 8.5cm
\epsfbox{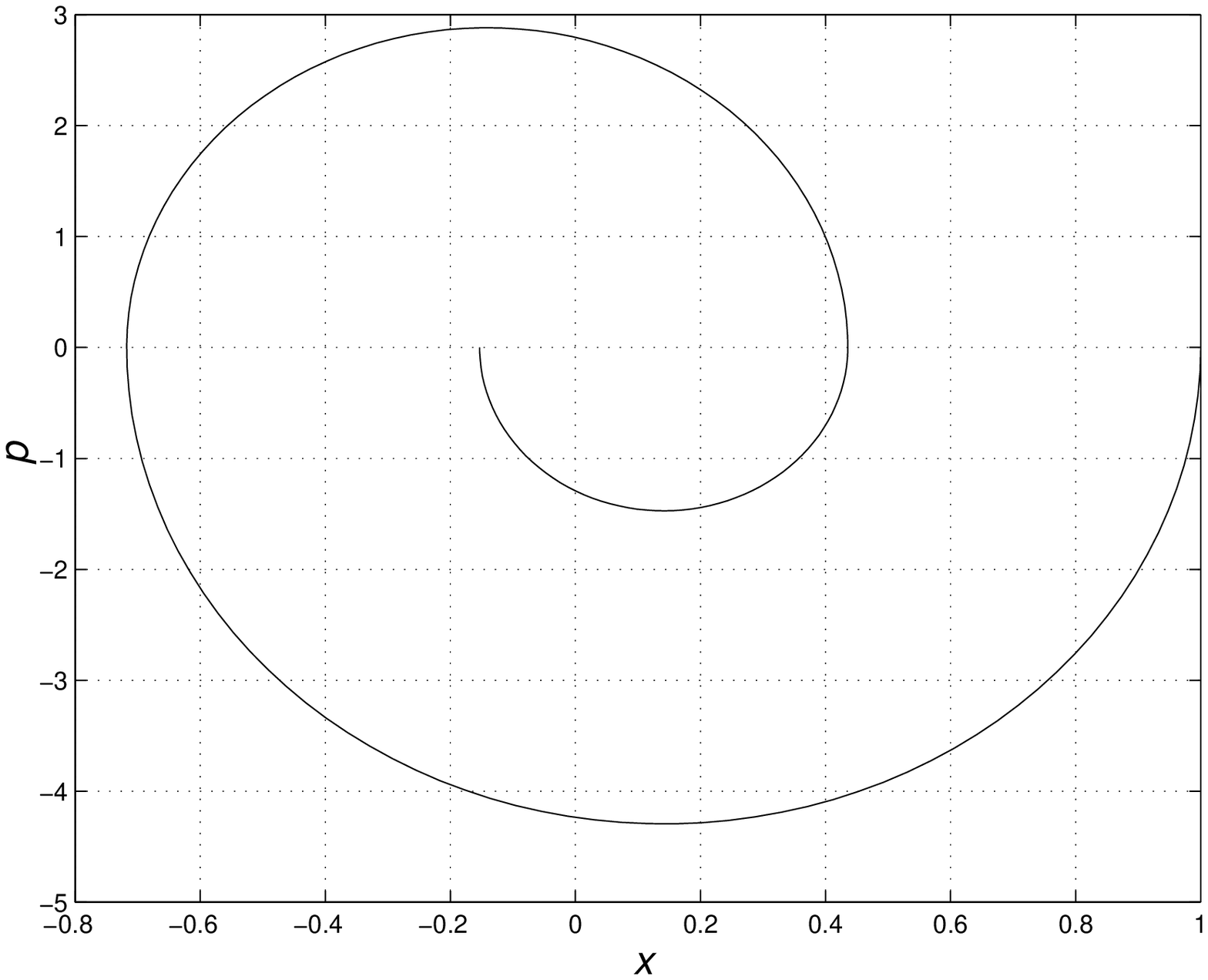}}
\caption{The orbit of the system in phase space: the momentum
$p=M\dot{x}$ vs. the position $x$, for the
choice of parameters $\mu_s=0.54$ and $\mu_k=0.36$.}
\label{fig3} \end{figure}

\begin{figure}
\input epsf
\centerline{ \epsfxsize 8.5cm
\epsfbox{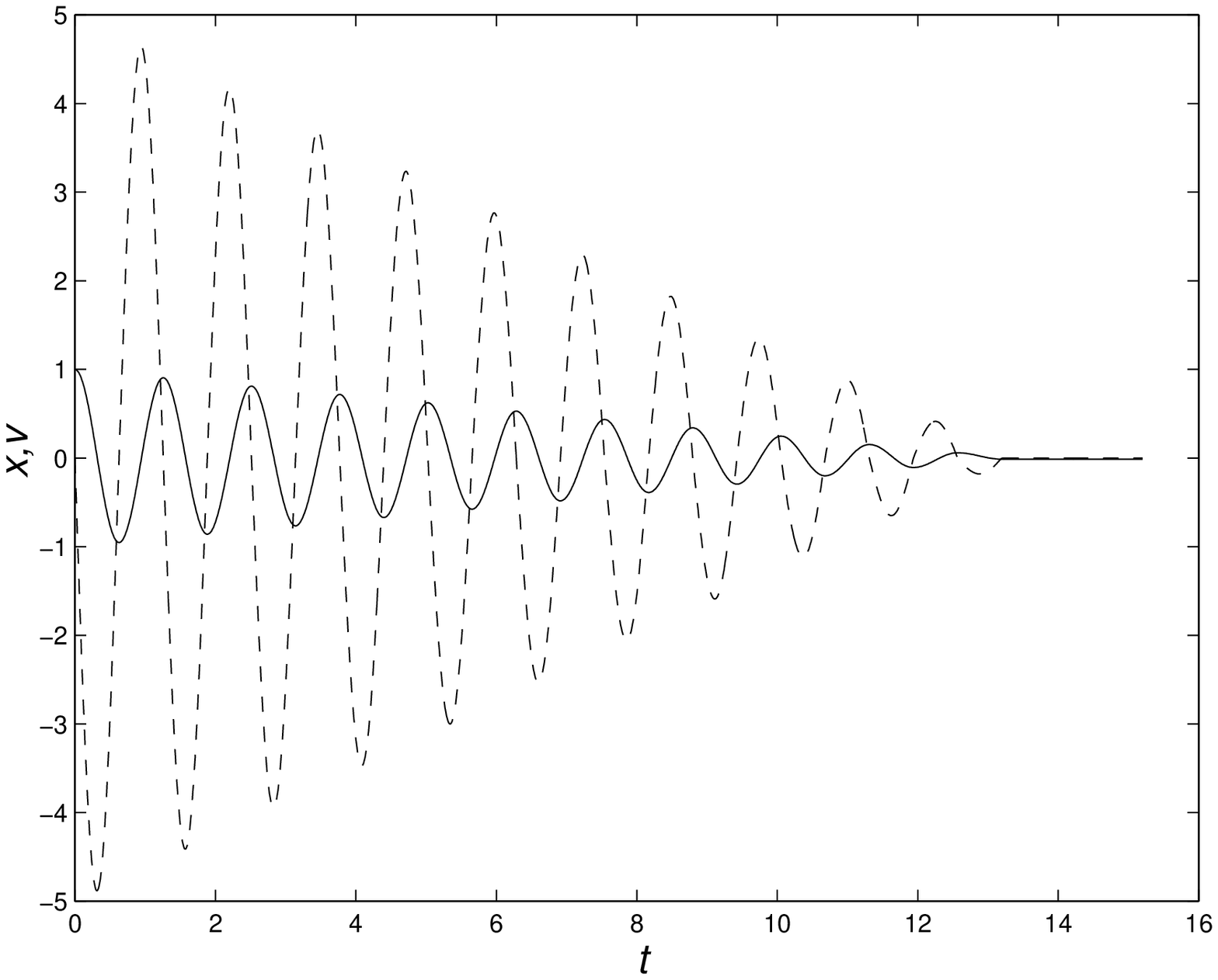}}
\caption{The position and the velocity as functions of time $t$, for the
choice of parameters $\mu_{s}=0.15$ and $\mu_{k}=0.06$. Solid line: $x(t)$.
Dashed line: $v\equiv\dot{x}(t)$.}  
\label{fig4} \end{figure}

\begin{figure}
\input epsf
\centerline{ \epsfxsize 8.5cm
\epsfbox{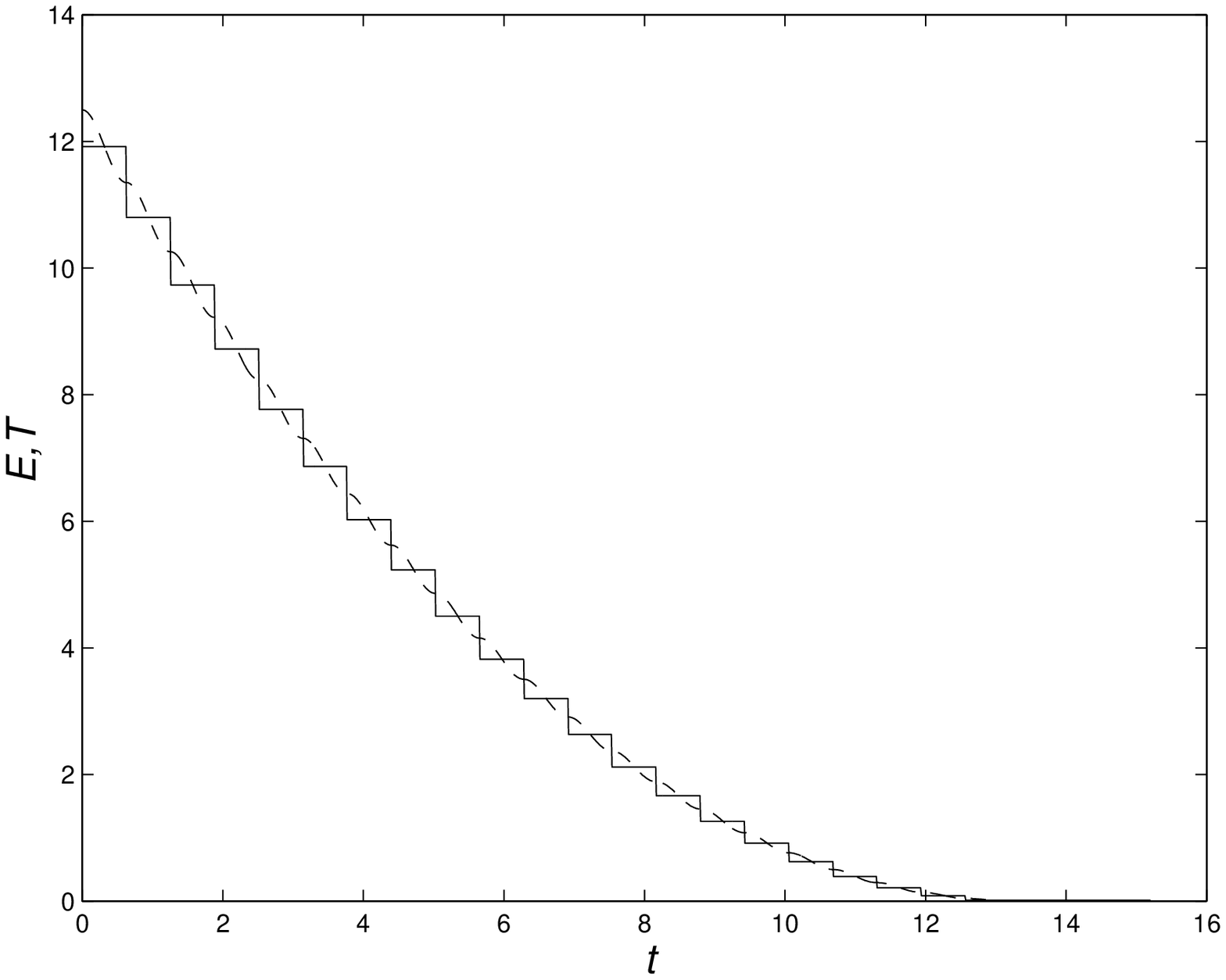}}
\caption{The picewise-conserved constant of motion and the total
mechanical energy as functions of time $t$, for the
choice of parameters $\mu_{s}=0.15$ and $\mu_{k}=0.06$. Solid line: ${\cal
E}(t)$.   Dashed line: $T(t)$.}
\label{fig5} \end{figure}

\begin{figure}
\input epsf
\centerline{ \epsfxsize 8.5cm
\epsfbox{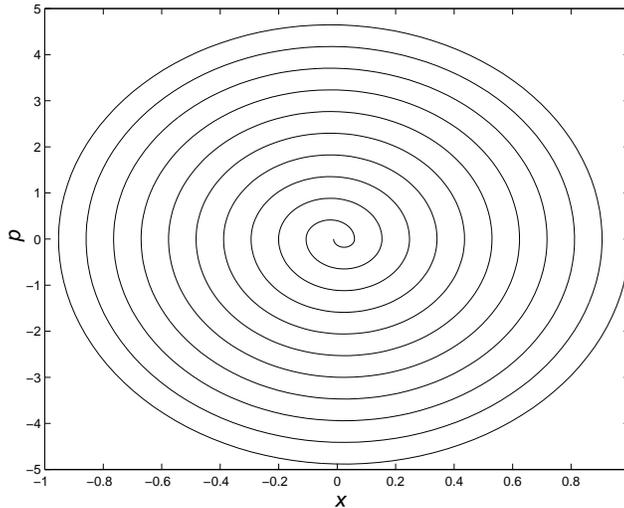}}
\caption{The orbit of the system in phase space: the momentum
$p=M\dot{x}$ vs. the position $x$, for the 
choice of parameters $\mu_{s}=0.15$ and $\mu_{k}=0.06$.}
\label{fig6} \end{figure}

We next present our results graphically for two sets of parameters.
First, we choose the parameters $\ell=1\,{\rm m}$, $\omega=5\,{\rm
sec}^{-1}$, $M=1\,{\rm Kg}$, $g=9.8\,{\rm
m}/{\rm sec}^2$, $\mu_s=0.54$, and $\mu_k=0.36$. (These values for the
coefficients of friction are typical for copper on steel.) In all the
figures below the units of all axes are SI units. Figure \ref{fig1} displays
the position $x(t)$ and the velocity
$\dot{x}(t)$ vs. the time $t$. It is
clear that the
amplitude of the oscillation attenuates, and eventually the block stops
in a state of rest. Figure \ref{fig2} displays the piecewise-conserved
${\cal E}$ and the mechanical energy $T$ as functions of the time $t$.
Indeed, the energy $T(t)$ is a monotonically-decreasing function of $t$,
whereas ${\cal E}(t)$ is piecewise-conserved. One can also observe that
up to a constant indeed ${\cal E}$ is the average of the energy $T$ over one
half-cycle of the motion.The dissipation of energy 
is most clearly portrayed by means of the phase space. Figure \ref{fig3}
shows the orbit of the system in phase space, namely, the momentum
$p=M\dot{x}$ vs.
the position $x$. The loss of energy is evident from the inspiral of the
orbit. Eventually, the orbit arrives at a final position in phase space,
and stays there forever. For figures \ref{fig4},\ref{fig5}, and
\ref{fig6} we changed only the coefficients of friction to
$\mu_s=0.15$ and $\mu_k=0.06$. (These parameters are typical for the
contact of two lubricated metal surfaces.)
As the coefficients of friction in this case are
smaller than their counterparts in the former case, we can observe many
more cycles of motion before the motion stops. (In fact, the number of
half-cycles in this case agrees with Eq. (\ref{eq20}) below.)  We note that
because of
the scale of Fig. \ref{fig5} it is not apparent that the energy arrives
at a non-zero constant value at late times. In this case, also, the
qualitative characteristics of the motion are the same as in the former
case (Figs. \ref{fig1}--\ref{fig3}), but here the attenuated oscillatory
motion is more apparent.

Of course, the motion will not continue forever: Because of the decrease 
in the amplitude of the motion, eventually the static friction force 
at some turning point would be larger than the elastic force exerted on 
the block by the spring. Namely, at $t=nP_{1/2}$ we find 
$x(t=nP_{1/2})=(-1)^n(\ell -2n\mu_kg/\omega^2)$, for some integer $n$, and 
the motion will stop 
for $\mu_sg\ge\omega^2(\ell -2n\mu_kg/\omega^2)$, or after an integral 
number of phases which is equal to the least integer $n$ which satisfies 
\begin{equation}
n\ge\frac{1}{2}\left(\frac{\omega^2\ell}{\mu_kg}-
\frac{\mu_s}{\mu_k}\right).
\label{eq20}
\end{equation}
We note that for the special case where $\omega^2\ell^2/(\mu_kg)$ is 
integral the block may stop at $x=0$. This happens, however, only for 
special values of the parameters of the systems, and in general the 
system will rest at $x\ne 0$.

Then, the block would remain at rest, and ${\cal 
E}$ would be a constant of motion from then on. Namely, because of the 
dissipative nature of the problem, eventually the piecewise-conserved 
constant of motion becomes a true constant of motion, but this happens 
only when the dynamics of the system becomes trivial. (In our case, when 
the system is in a constant state of rest.) This feature of the 
dissipative system is in contrast with other piecewise-conserved 
constants of motion, which arise from non-dissipative systems, such as 
the truncated Kepler problem or the three-dimensional isotropic harmonic 
oscillator, where the piecewise-conserved constant vector remains 
piecewise conserved for all times.

\end{document}